\documentclass[letterpaper,twocolumn,showpacs,
prb
,superscriptaddress,email
]{revtex4}

\usepackage[pdftex]{graphicx}  % needed for figures
\usepackage{amssymb}
\usepackage{amsmath,amsfonts,latexsym}
\usepackage{array,tabularx,color}
\usepackage{dcolumn}               % Align table columns on decimal point
\usepackage[normalem]{ulem}
\usepackage{comment}
\usepackage{bm}
\usepackage[latin1]{inputenc}
\usepackage{color}

\newcommand{\vect}[1]{{\mathbf #1}}

\begin{document}

\title{Multistability of a two component exciton-polariton fluid}

\author{E.~Cancellieri}
\email{emiliano.cancellieri@uam.es}
\affiliation{F\'{\i}sica Te\'orica de la Materia Condensada,
  Universidad Aut\'onoma de Madrid, Madrid 28049, Spain.}

\author{F. M. Marchetti}
\affiliation{F\'{\i}sica Te\'orica de la Materia Condensada,
  Universidad Aut\'onoma de Madrid, Madrid 28049, Spain.}

\author{M. H. Szyma\'nska}
\affiliation{Department of Physics, University of Warwick, Coventry,
  England.}
\altaffiliation{also at London Centre for Nanotechnology, UK}

\author{C. Tejedor}
\affiliation{F\'{\i}sica Te\'orica de la Materia Condensada,
  Universidad Aut\'onoma de Madrid, Madrid 28049, Spain.}

%\date{\today}
\date{February 15, 2012}       % to fix in the last version!

\begin{abstract}
  We study the stability of a multicomponent exciton-polariton fluid
  under resonant excitation within the linear response approximation
  of a generalized Gross-Pitaevskii equation. We show that, two
  spatially homogeneous and independently tunable pumping lasers
  produce, for the same values of the system parameters, up to three
  stable solutions.  Three-stability is understood by noting that the
  cavity can be either little or highly populated and, in this second
  case, the largest part of the population lies in either one of the
  two components. Moreover, we discuss the different kinds of
  instabilities appearing at different pumps intensities and compare
  them with the case of one-component fluids.  Finally, we show that
  easily tunable multistable hysteresis loops can be performed by the
  system.
\end{abstract}

%03.75.Kk   Dynamic properties of condensates; collective and
%           hydrodynamic excitations, superfluid flow
%71.36.+c   Polaritons (including photon-phonon and
%           photon-magnon interactions)
%41.60.Bq   Cherenkov radiation
%42.65.Pc   Optical bistability, multistability, and switching,
%           including local field effects 
%           (see also 42.60.Gd Q-switching; 42.79.Ta Optical
%           computers, logic elements, interconnects, switches; neural
%           networks)
\maketitle

\section{Introduction}
Condensates of resonantly pumped exciton-polaritons in semiconductor
microcavities constitute a novel and exciting system for the study of
fundamental physical properties of superfluids out of
equilibrium~\cite{keeling09}, and for future device
applications~\cite{menon10,amo10}.  Being quantum superpositions of
light and matter they are privileged candidates for the realization of
the next generation of optical devices~\cite{menon10}, for example for
quantum information technologies.

From the point of view of the fundamental physical properties,
particularly interesting is the case of a coherently pumped polariton
superfluid in presence of defects. Here, in contrast to the
corresponding equilibrium case, a weak residual drag force is always
present even at extremely high polariton
densities~\cite{cancellieri10}. Nevertheless, properties
paradigmatical of an equilibrium superfluids, such as frictionless
flow of polariton bullets~\cite{amo09}, quantized vortices and
metastable persistent flow~\cite{sanvitto09}, and the appearance and
disappearance of {\u C}erenkov-like waves~\cite{amo09_b} have been
recently observed in coherently driven exciton-polaritons.

In view of the potential device applications, especially important is
the unique versatility of the polaritonic system, which, combined with
its high non-linear properties, have been already demonstrated to
produce parametric
scattering~\cite{carusotto04,savvidis00:prl,stevenson00} and
bistability~\cite{baas04}. Here, the implementation of logic
operations and gates comes in a natural way: By manipulating the
non-linear properties of the system using several lasers which
frequencies, angles of incidence, and intensities can be freely varied
externally.

In this paper we investigate a new realization of a two-component
polariton system coherently driven by two lasers with independently
tunable frequencies, angles of incidence and intensities.  Firstly, we
study the stability of the two polariton components when the two laser
intensities are varied.
Differently from the case of a single laser pump, where the system can
only be bistable, we disclose a rich phase diagram, where either one,
two or three stable states can coexist at given pumping conditions.
Then we suggest possible easily tunable multistable hysteresis cycles
when the two pumping lasers are varied up and down in intensities.

Alternatively, a multicomponent polariton fluid can be realized by
considering the polarization degrees of freedom. Multistability of
different polariton spin states has been recently proposed
theoretically~\cite{gippius07} and confirmed
experimentally~\cite{paraiso10} by the observation of three stable
spin states for a given excitation condition.  For the case of two
polarized components, multistability in space have also been
theoretically proposed~\cite{shelykh08,shelykh06} and experimentally
observed~\cite{adrados10}.  For the system with two pumping lasers
presented here, the same kind of spatial multistability is expected
but with much more complex features.  Additionally, superimposed to
the spatial multistability, interference fringes will appear due to
the difference in frequency and momentum of the two pumping lasers.
The analogy with the two-component polarized case, suggests that,
aside from the interest in investigating multistability, two-component
polariton condensates obtained with independent lasers can also be
used to realize switches~\cite{amo10} and memories.

The paper is organized as follows: in section II, we present the model
used to describe the steady state behavior of polaritons excited by
two continuous-wave lasers with different frequencies, wave-vectors
and intensities. The results obtained within a linear response
framework are shown in section III. In this section we study the
number of the possible solutions, their nature and possible Kerr or
parametric instabilities associated to them. As a consequence of the
presence of multiple stable solutions, different cycles of hysteresis
can be produced by varying, along different paths, the intensities of
the two pumping lasers. Finally, section IV contains the conclusions
drawn from our analysis.

\section{Model}
The dynamics of resonantly-driven microcavity
polaritons~\cite{ciuti03,carusotto04} can be described via a
Gross-Pitaevskii equation for coupled cavity ($\psi_C$) and exciton
($\psi_X$) fields generalized to include decay and resonant pumping
($\hbar=1$):
\begin{equation}
  i\partial_t \begin{pmatrix} \psi_X \\ \psi_C \end{pmatrix} =
  \begin{pmatrix} 0 \\ F \end{pmatrix} +
  \left[\hat{H}_0 + \begin{pmatrix} g_X|\psi_X|^2& 0 \\ 0 &
      0 \end{pmatrix}\right]
  \begin{pmatrix} \psi_X \\ \psi_C \end{pmatrix}\; .
\label{eq:model}
\end{equation}
The repulsive ($g_X>0$) exciton-exciton interaction induces a
non-linear dynamics of the eigenmodes of the single polariton
Hamiltonian (lower and upper polariton, $\omega_{LP,UP}(\vect{k})$):
\begin{equation}
  \hat{H}_0 = \begin{pmatrix} \omega_{X} (-i\nabla) - i \kappa_X &
    \Omega_R/2 \\ \Omega_R/2 & \omega_{C}(-i\nabla) - i
    \kappa_C \end{pmatrix} \; .
\end{equation}
Here, we assume the cavity dispersion to be quadratic,
$\omega_C(\vect{k})=\omega_C(0)+ k^2/(2m_C)$, with
$m_C=2\times10^{-5}m_0$ ($m_0$ is the bare electron mass), we will
neglect the exciton dispersion and consider the case of zero detuning
at normal incidence, $\omega_X(\vect{k}) = \omega_X
(0)=\omega_C(0)$. The Rabi frequency $\Omega_R=5.0$~$[$meV$]$ and the
excitonic and photonic decay rates, $\kappa_X=\kappa_C=0.05$~$[$meV$]$
are chosen in the range of experimental values.

Because of the continuous decay, a stationary state requires a
continuous injection of photons. Here, we consider two continuous-wave
laser fields,
\begin{equation}
  F(\vect{r},t)=F_1 e^{i (\vect{k}_1 \cdot \vect{r} - \omega_1 t)}
  +F_2 e^{i (\vect{k}_2 \cdot \vect{r} - \omega_2 t)}\; ,
\end{equation}
with independently tunable frequencies $\omega_{1,2}$ and momenta
$\vect{k}_{1,2}$, which can be experimentally changed by changing the
laser angle of incidence with respect to the growth direction.

We study the mean-field solutions of Eq.~\eqref{eq:model}
\begin{equation}
  \psi_{X,C}(\vect{r},t) =
  \psi^{ss}_{1_{X,C}}e^{i(\vect{k}_1\cdot\vect{r}-\omega_1 t)} +
  \psi^{ss}_{2_{X,C}}e^{i(\vect{k}_2\cdot\vect{r}-\omega_2 t)} \; ,
\label{eq:meanf}
\end{equation}
and their stability with respect to small fluctuations within a linear
response analysis. Substituting the expression~\eqref{eq:meanf}
into~\eqref{eq:model} we obtain 4 contributions, two of which
oscillate at the main frequencies $\omega_1$ and $\omega_2$ and the
additional two at the replica (or satellite state) frequencies
$\omega_1-\Delta\omega$ and $\omega_2+\Delta\omega$, where $\Delta
\omega=\omega_2 - \omega_1$. Similarly to what is done in the OPO
regime~\cite{whittaker05,wouters07} where replica states in addition
to the pump signal and idler states are neglected, here, we consider
only the terms oscillating at the main frequencies $\omega_1$ and
$\omega_2$. Later, see Eq.~\eqref{eq:fluct}, we analyse the dynamical
stability of the two-pump-frequency solution against the weak
population of satellite states $\omega_i \pm \omega$ via parametric
scattering processes. Through the paper, we will consider only
dynamically stable two-pump-frequency solutions.
In this approximation, we obtain the following mean-field equations
for $\psi^{ss}_{{1,2}_{X,C}}$:
\begin{equation}
  \begin{cases}
  [\omega_X-\omega_1-i\kappa_X+G_{12}]\psi_{1_X}^{ss}+
    \frac{\Omega_R}{2}\psi_{1_C}^{ss}=0 \\
  [\omega_C({\bf k}_1)-\omega_1-i\kappa_C]\psi_{1_C}^{ss}+
  \frac{\Omega_R}{2}\psi_{1_X}^{ss}+F_1 =0 \\
  [\omega_X-\omega_2-i\kappa_X+G_{21}]\psi_{2_X}^{ss}+
  \frac{\Omega_R}{2}\psi_{2_C}^{ss}=0 \\
  [\omega_C({\bf k}_2)-\omega_2-i\kappa_C]\psi_{2_C}^{ss}+
  \frac{\Omega_R}{2}\psi_{2_X}^{ss} +F_2 =0 \; ,
\end{cases}
\label{eq:stead}
\end{equation}
where $G_{ij}=g_X(|\psi_{i_X}^{ss}|^2+2|\psi_{j_X}^{ss}|^2)$ with
$i\neq j=1, 2$.
Note that the repulsive interaction term between excitons in different
states is two times larger the interaction term between excitons in
the same mode, resulting in a non-uniform blue-shift.  The mean-field
system of equations~\eqref{eq:stead} can have up to 9 solutions,
i.e. 6 solutions more than in the case of one pumping laser, but, as
discussed below, only a maximum of 3 solutions are stable.

The dynamical stability of the two-pump-frequency mean-field solution
can be established by adding small fluctuations,
\begin{multline}  
  \psi_{X,C}(\vect{r},t) = e^{-i\omega_1 t}
  \left[e^{i\vect{k}_1\cdot\vect{r}} \psi^{ss}_{1_{X,C}} +
    \theta_{1_{X,C}}(\vect{r,t})\right] +\\
  e^{-i\omega_2 t}
  \left[e^{i\vect{k}_2\cdot\vect{r}} \psi^{ss}_{2_{X,C}} +
    \theta_{2_{X,C}}(\vect{r,t})\right] \; ,
\label{eq:fluct}
\end{multline}
where the fluctuation fields can be divided into particle-like and
hole-like excitations $\theta_{i_{X,C}} (\vect{r},t) = \sum_{\vect{k}}
[e^{-i\omega t + i\vect{k}\cdot \vect{r}} u_{i_{X,C}\vect{k}} +
  e^{i\omega t + i(2 \vect{k}_i -\vect{k})\cdot \vect{r}}
  v^*_{i_{X,C}\vect{k}}]$. Expanding Eq.~\eqref{eq:model} up to linear
terms in $\theta_{1,2_{X,C}}$, we obtain 4 terms oscillating at
frequencies $\omega_1-\Delta\omega \pm \omega$ and
$\omega_2+\Delta\omega \pm \omega$, which we neglect, and 4 terms
oscillating at $\omega_i \pm \omega$. In other words, we are checking
the stability of our solution, where only the two states with
frequencies $\omega_{1,2}$ are occupied, against the weak population
of the satellite states $\omega_i \pm \omega$ which can be populated
by parametric scattering processes. The fact that we consider only
linear terms in $u_{i_{X,C}\vect{k}}$ and $v^*_{i_{X,C}\vect{k}}$
implies that we can obtain only the threshold conditions for such
parametric processes, as well as the nature of the instability,
whether of Kerr-type or parametric-type --- see later. The equations
for $u_{i_{X,C}\vect{k}}$ and $v^*_{i_{X,C}\vect{k}}$ can be written
as an eigenvalue equation rearranging the excitations into an
8-component vector $\mathbb{U}^{\text{T}} =
(u_{1_X},u_{1_C},v_{1_X},v_{1_C},u_{2_X},u_{2_C},v_{2_X},v_{2_C})$:
\begin{equation}
  \left[\omega \mathbb{I} - \begin{pmatrix} \mathbb{L}_{11 \vect{k}}&
      \mathbb{L}_{12 \vect{k}} \\ \mathbb{L}_{21 \vect{k}} &
      \mathbb{L}_{22 \vect{k}} \end{pmatrix}\right]
  \mathbb{U}_{\vect{k}} = 0\; .
\label{eq:eigen}
\end{equation}
Here matrices $\mathbb{L}_{ij \vect{k}}$ with $i\ne j$ are given by
\begin{equation*}
  2g_X e^{i(\vect{k}_i - \vect{k}_j)\cdot \vect{r}} \begin{pmatrix}
    \psi_{i_X}^{ss}\psi_{j_X}^{ss\star} & 0 & \psi_{i_X}^{ss}
    \psi_{j_X}^{ss} & 0 \\ 0 & 0 & 0 & 0 \\ -
    \psi_{i_X}^{ss\star}\psi_{j_X}^{ss\star} & 0 & -
    \psi_{i_X}^{ss\star} \psi_{j_X}^{ss} & 0 \\ 0 & 0 & 0 & 0
  \end{pmatrix}
\end{equation*}
and $\mathbb{L}_{jj \vect{k}}$ are given by
\begin{widetext}
\begin{equation*}
  \begin{pmatrix}
  \omega_X - \omega_j - i\kappa_X + g_X|\psi_X^{ss}|^2 &
  \frac{\Omega_R}{2} & g_X\psi_{j_X}^{ss}\psi_{j_X}^{ss} & 0
  \\ \frac{\Omega_R}{2} & \omega_C(\vect{k}) - \omega_j - i\kappa_C &
  0 & 0 \\ -g_X \psi_{j_X}^{ss\star} \psi_{j_X}^{ss\star} & 0 &
  -\omega_X (2 \vect{k}_j- \vect{k}) + \omega_j - i\kappa_X -
  g_X|\psi_X^{ss}|^2 & -\frac{\Omega_R}{2} \\ 0 & 0 &
  -\frac{\Omega_R}{2} & - \omega_C (2 \vect{k}_j- \vect{k}) + \omega_j
  - i\kappa_C \end{pmatrix}\; ,
\end{equation*}
\end{widetext}
with $|\psi_X^{ss}|^2=2(|\psi_{1_X}^{ss}|^2+|\psi_{2_X}^{ss}|^2)$
being the total excitonic density. At given values of the pumping
strength $F_1$ and $F_2$, the solutions of the mean-field
equations~\eqref{eq:stead} are stable if all the eight eigenvalues
($\text{LP}_j^{\pm}(\vect{k})$, $\text{UP}_j^{\pm}(\vect{k})$) of
Eq.~\eqref{eq:eigen} have negative imaginary part for every value of
the momentum $\vect{k}$.

\section{Results}
For some choices of the system parameters, we find that the number of
stable solutions can be larger than one. In the case of one pumping
laser, the typical $S$-like shape dependence of the polariton field
intensity on the pump strength, also referred to as optical
bistability, can be explained in terms of the non-linear blue-shift
induced by the polariton-polariton
interaction~\cite{whittaker05,ciuti05}.  When the laser frequency is
well above the bare lower polariton dispersion, $\omega_p >
\omega_{LP} (\vect{k}_p)$, and the pump intensity increases from low
values, the polariton population remains small because it is hard for
the laser to inject polartions with a different energy. However,
increasing the pump power, the blue-shift pulls the polariton energy
towards resonance with the pump causing the population to grow
superlinearly and eventually to abruptly jump to a high value when the
pump intensity reaches a critical value $I_1$. In the opposite
situation, when the laser intensity is decreased from high values, the
polariton energy is blue-detuned close to the pumping laser frequency
and, therefore, the cavity is efficiently filled by the laser even at
low pumping intensities. In this case the polariton population jumps
down back at low polariton densities for a value $I_2$ of the pump
strength lower than $I_1$. The two jumps at different values of the
pump intensity cause therefore a hysteresis cycle. As explained below,
in the case of two-component fluids, the situation becomes even
richer.
\begin{figure}
\begin{center}
\includegraphics[width=1.0\linewidth,angle=0]{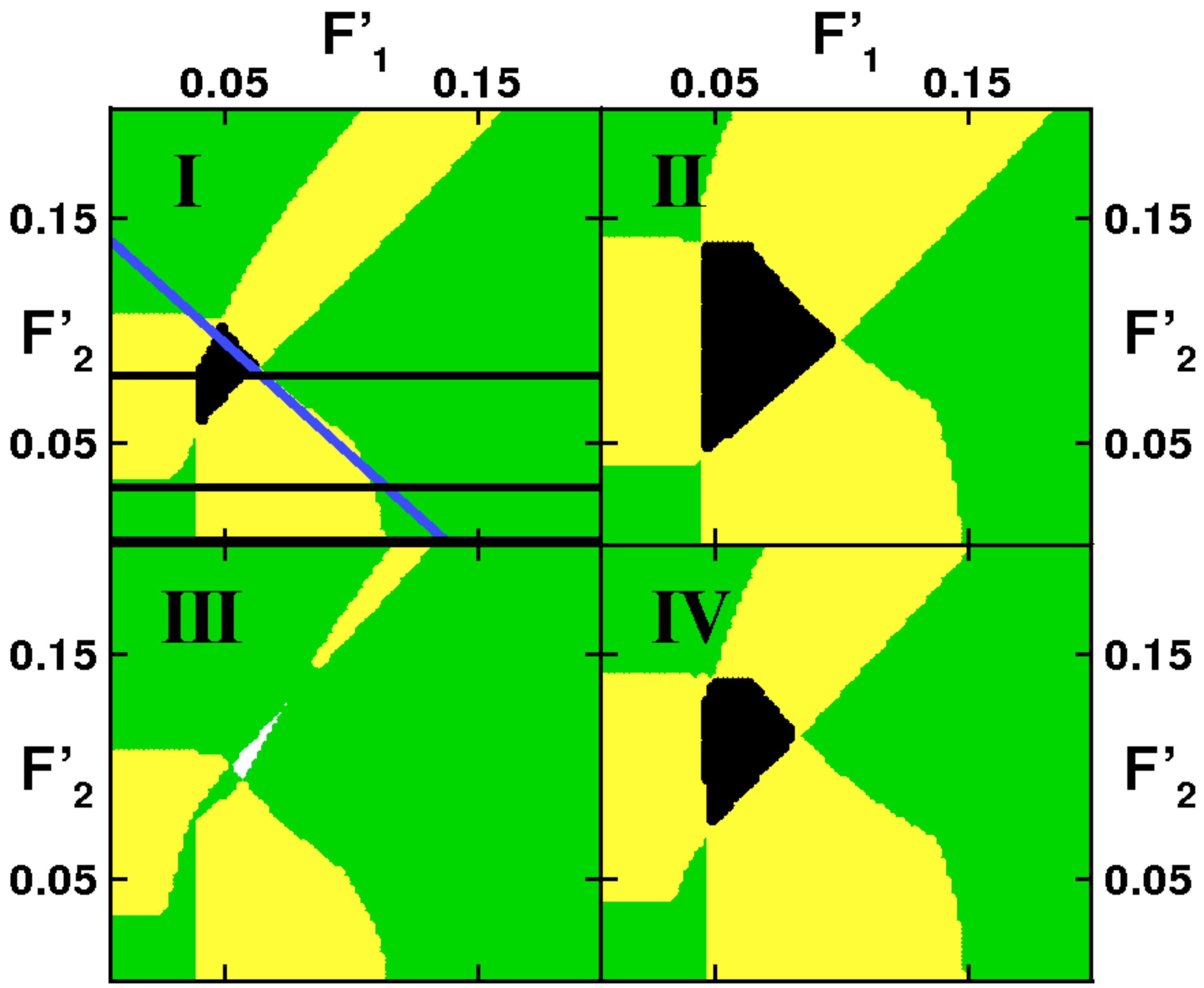}
\includegraphics[width=1.0\linewidth,angle=0]{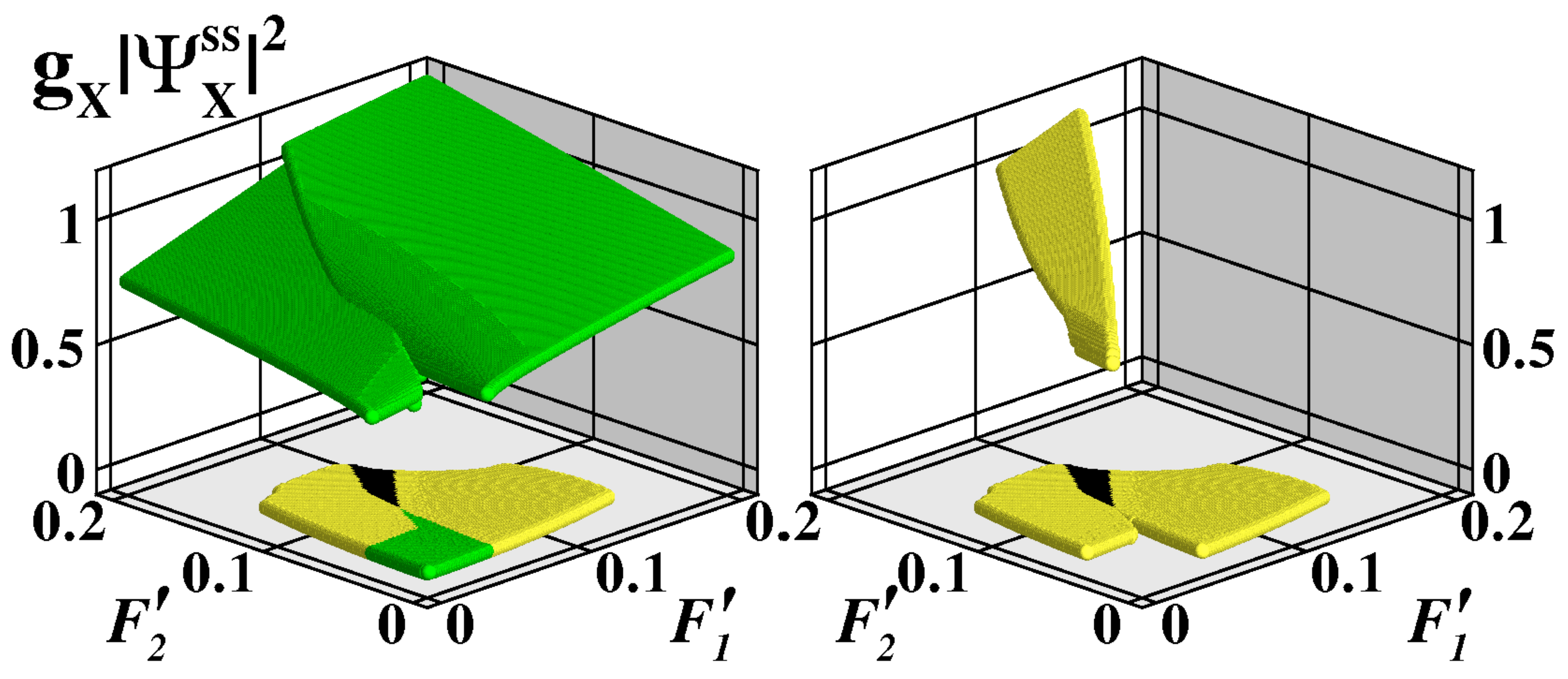}
\end{center}
\caption{(Color online) 2D panels: phase diagram showing the number of
  stable solutions as a function of the rescaled pump intensities
  $F_{1,2}'=\sqrt{g_x} F_{1,2}$ [meV$^{3/2}$]. White, green, yellow
  and black regions correspond to respectively zero, one, two or three
  stable solutions. In panels I and II $\vect{k}_1=0.25$~$\mu$m$^{-1}$
  and $\vect{k}_2=0.7$~$\mu$m$^{-1}$ while in panels III and IV
  $\vect{k}_1=0.0$~$\mu$m$^{-1}$ and
  $\vect{k}_2=0.7$~$\mu$m$^{-1}$. In the left panels (I and III)
  $\omega_{1,2}=\omega_{LP} (\vect{k}_{1,2})+0.3$~$[$meV$]$, while in
  the right panels (II and IV) $\omega_{1,2}=\omega_{LP}
  (\vect{k}_{1,2})+0.4$~$[$meV$]$. The horizontal black lines lies at
  the three fixed values of $F_2'$ corresponding to the three panels
  of Fig.~\ref{fig:cuts}, while the blue diagonal line is the path
  used to plot Fig.~\ref{fig:hyst}. 3D panels: plots of
  $g_X|\psi_{X}^{ss}|^2$ [meV] as a function of $F_{1,2}'$ with
  parameters equal to panel I. Stable solutions with higher
  populations are shown in green, stable solution with the second
  higher population in yellow and third stable solution with lower
  population in black. All the solutions are shown in the left
  panel. Since the upper green branches hide a yellow upper branch,
  the right panel shows only the yellow and black solutions.}
\label{fig:2d3d}
\end{figure}

\begin{figure}
\begin{center}
\includegraphics[width=1.0\linewidth,angle=0]{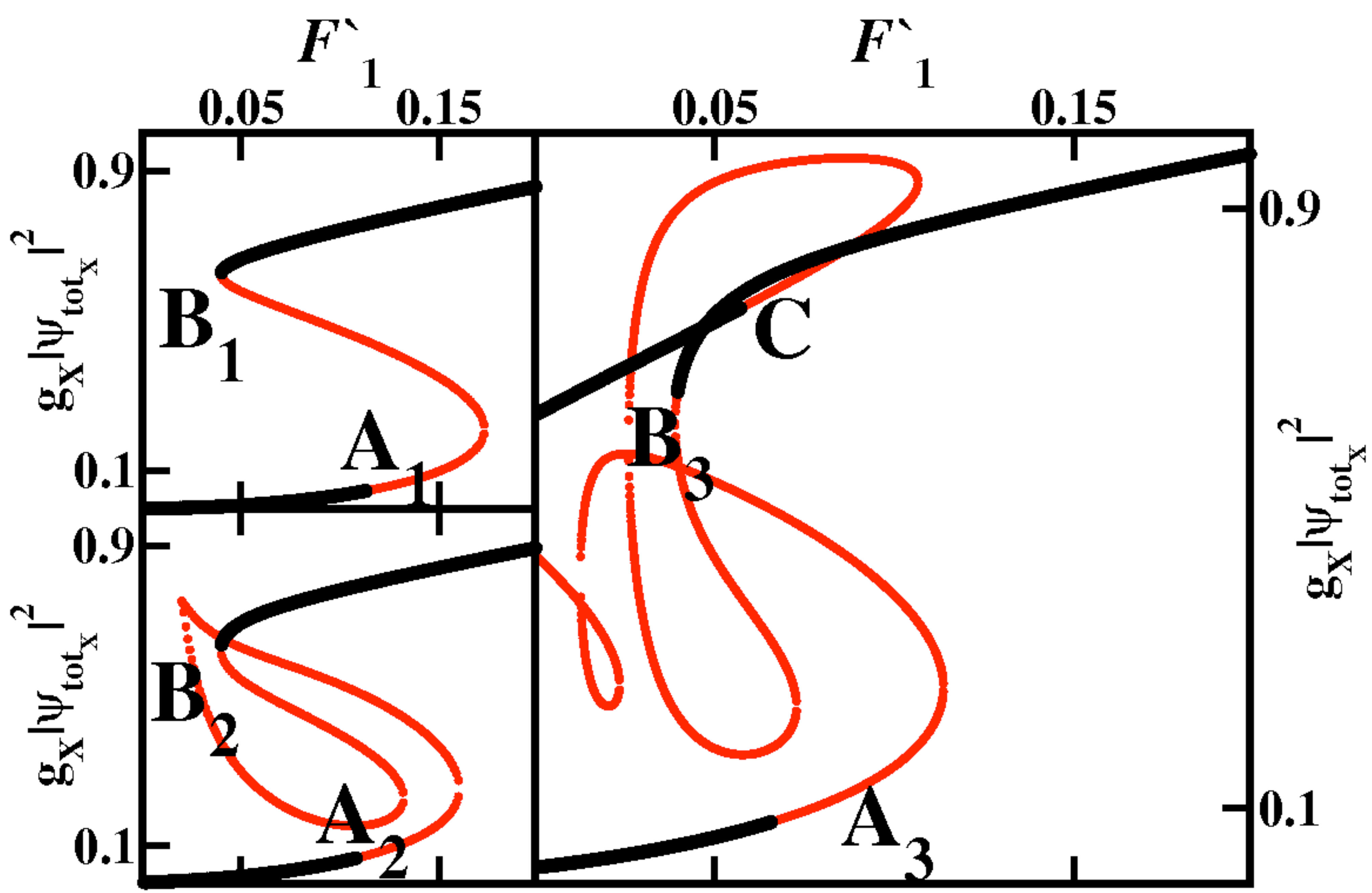}
\end{center}
\caption{(Color online) Stability curves of the total exciton density
  $g_X |\psi_X^{ss}|^2$ $[$meV$]$ (red dotted curves unstable
  solutions, black lines stable solutions) for fixed pump intensities
  as a function of $F_1'$, for $F_2'=0.00001$ $[$meV$^{3/2]}$ (top
  left), $F_2'=0.025$ $[meV^{3/2}]$ (bottom left), and $F_2'=0.08$
  $[$meV$^{3/2}]$ (right).  Points $A_i$ ($B_i$), $i=1,2,3$ correspond
  to the cases when the lower (upper) branch of the stability curve
  becomes unstable (see Fig.~\ref{fig:imaginary1} and
  ~\ref{fig:imaginary2}).  Point $C$ does not have a counterpart in
  the one-fluid case and correspond to the cases when the second high
  branch of the stability curve becomes unstable (see
  Fig.~\ref{fig:imaginary2}).}
\label{fig:cuts}
\end{figure}

We fix both laser frequencies to be blue detuned with respect to the
bare polariton dispersion: $\omega_{1,2}=\omega_{LP}
(\vect{k}_{1,2})+0.3$~$[$meV$]$, with $\vect{k}_1=0.25$~$\mu$m$^{-1}$
and $\vect{k}_2=0.7$~$\mu$m$^{-1}$. We plot in panel I of
Fig.~\ref{fig:2d3d} the phase diagram showing the regions with a
different number of stable solutions (either one, two or three) as a
function of the two rescaled pumping intensities $F_{1,2}'=\sqrt{g_x}
F_{1,2}$ [meV$^{3/2}$]. In order to understand better the structure of
this phase diagram, we show in Fig.~\ref{fig:cuts} the total exciton
density $g_X |\psi_X^{ss}|^2$ when the pump intensities $F_2'$ is kept
constant at different values and $F_1'$ is varied. When the constant
pump $F'_2$ has a small value (Fig.~\ref{fig:cuts} top left panel),
the dependence of the population on the varying pump intensity $F_1'$
is similar to the one-fluid case showing bistability with an $S$-like
shape. For higher values of the constant pump $F_2'$ (Fig.~\ref{fig:cuts}
lower left panel), the number of possible solutions increases but just
two are found to be stable. Finally when $F_2'$ is further increased
the set of possible solutions further goes up but only a maximum of
$3$ are found to be stable.

The coexistence of three solutions, corresponding to the black regions
of Fig.~\ref{fig:2d3d}, can be understood as follows: when the two
pump intensities increase from low values, the polariton population is
small because it's energy is far below the laser frequencies,
resulting in one stable solution. In the opposite situation, when the
intensity of one of the lasers decreases from high values, the
polariton population is high and its dispersion is significantly blue
detuned with respect to the bare one. Such blue-shift can be sustained
by any of the two lasers, thus giving two additional stable solutions
for the same values of the pump intensities. Therefore a maximum of
three stable solutions can be expected. This is also evident while
considering the partial densities for particular values of the pump
strength at which three stable solutions are present (black region of
Fig.~\ref{fig:2d3d}), e.g. $F_1'\equiv \sqrt{g_X} F_1= 0.05$
$[$meV$^{3/2}]$ and $F_2'\equiv \sqrt{g_X} F_1= 0.08$
$[$meV$^{3/2}]$. Here, the solution with lower total polariton density
corresponds to partial densities $g_X|\psi_{1_X}^{ss}|^2=
0.009$~$[$meV$]$ and $g_X|\psi_{2_X}^{ss}|^2=0.023$~$[$meV$]$. The
other two solutions correspond to a high value of just one of the two
partial populations: $g_X|\psi_{1_X}^{ss}|^2= 0.008$~$[$meV$]$ and
$g_X|\psi_{2_X}^{ss}|^2=0.609$~$[$meV$]$ in one case and
$g_X|\psi_{1_X}^{ss}|^2= 0.646$~$[$meV$]$ and
$g_X|\psi_{2_X}^{ss}|^2=0.010$~$[$meV$]$ in the other. Note that this
situation is similar to the case of two-component condensates obtained
with two spins. However, while in the spin-dependent case the two
lasers pump the two spin populations with different intensities but at
the same angle and energy, here the two pumps are independent also in
angle and in energy.
This analogy is also visible in Eq.~\eqref{eq:stead}. However, the
difference between our system of equations and the spin-dependent case
is that here the interaction between different components is twice the
interaction between particles in the same component.

In panel II of Fig.~\ref{fig:2d3d} we plot the phase diagram for the
same parameters as in panel I but with the two pumping lasers $0.4$
$[$meV$]$ blue detuned with respect to the bare LP branch. We see that
the effect of the increased detuning is simply to stretch the phase
diagram. Since the two pumps are further apart from the LP branch, it
is more difficult to inject polaritons into the cavity, and thus the
need for higher pump intensities.
In panel IV of Fig.~\ref{fig:2d3d} we show that a similar phase
diagram can be obtained by changing $\vect{k}$ vector of pump 1 from
$0.25$ to $0.0$~$\mu$m$^{-1}$. We observe that the multistability is
quite robust with respect to the choice of the parameters and,
therefore, it should be within an experimental reach. An interesting
configuration is plotted in panel III of the same figure. Here a
region with no stable solutions appears in the central part of the
plot (white region).  The instability of this region can be understood
by noting that with a pump at $\vect{k}=0.0$ and just slightly blue
detuned from the LP and a pump at $\vect{k}=0.7$, close to the
inflection point of the LP, it is easy to satisfy phase matching
conditions for parametric scattering processes. For this set of
parameters the system is in a configuration unstable towards the
population of satellite states by scattering processes. For the other
three sets of parameters, shown in the remaining three panels of
Fig.~\ref{fig:2d3d}, it is also possible to find regions of the phase
diagram where no solutions are stable. These are the regions where the
proposed solution, given by Eq.~\ref{eq:meanf}, where only the two
frequency states, $\omega_1$ and $\omega_2$, are populated, is not a
stable solution because satellite states start also to be populated
--- our analysis giving the threshold for this to happen.

\begin{figure}
\begin{center}
\includegraphics[width=1.0\linewidth,angle=0]{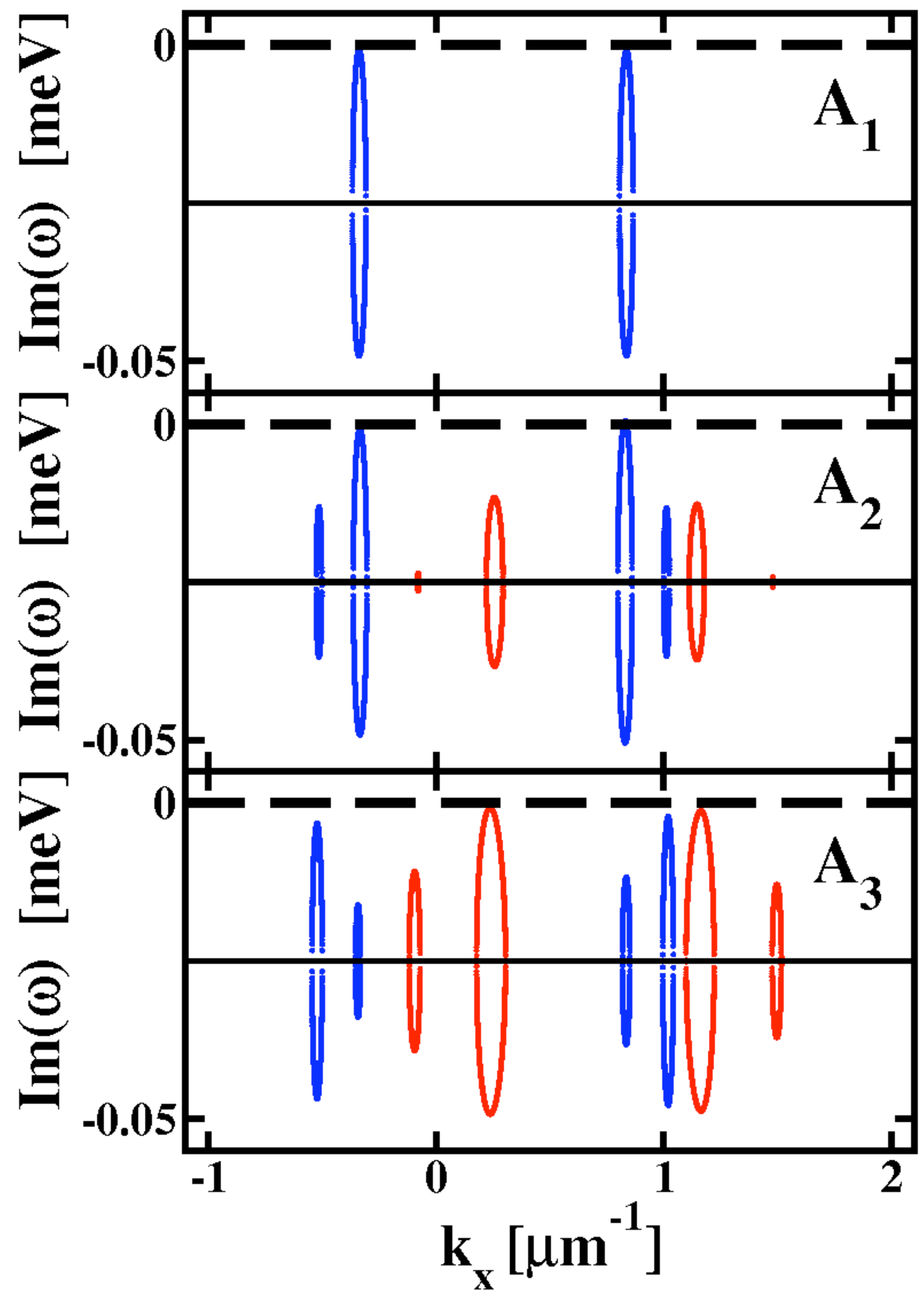}
\end{center}
\caption{(Color online) Dispersion of the imaginary part of the
  excitation eigenfrequency $\omega=\text{LP}_j^{\pm}$. The three
  panels correspond to points $A_i$ with $i=1,2,3$ of
  Fig.~\ref{fig:cuts}, where the lower part of the stability curves
  become unstable. In blue (red) the parts corresponding to the
  scattering of two particles with $\vect{k}=\vect{k}_1$
  ($\vect{k}=\vect{k}_2$).}
\label{fig:imaginary1}
\end{figure}

To further discuss the stability of the system with respect to small
perturbations, we plot the dispersion of the imaginary part of the
excitation eigenfrequency $\omega=\text{LP}_j^{\pm}$ for several
points of the stability curves shown in Fig.~\ref{fig:cuts}.  We start
with the cases where the lower branch of the stability curve became
unstable at points $A_i$. For very small values of $F_2'$ the
imaginary part of the dispersion (top panel of
Fig.~\ref{fig:imaginary1}) shows two peaks for given values of
$\vect{k}$. One peak lies at higher value $\vect{k}_+=0.83$
$\mu$m$^{-1}$ and one peak at lower value $\vect{k}_-=-0.33$
$\mu$m$^{-1}$ with $\vect{k}_++\vect{k}_-=2\vect{k}_1$. This two-peaks
structure is a precursor of a parametric instability due to the
scattering between two particles in the component of the condensate
with momentum $\vect{k}_1$. This situation corresponds exactly to the
case of one component fluids. When the pump intensity $F_2'$ is
slightly increased (middle panel) we observe 6 other peaks appearing
in the imaginary part of the dispersion. Two of these new peaks are
such that $\vect{k}_++\vect{k}_-=2\vect{k}_1$ (blue lines) while the
other four can be combined to identify two different scattering
processes with $\vect{k}_++\vect{k}_- =2\vect{k}_2$ (red curves). This
more complicated structure of the imaginary parts of the eigenvalues
is consistent with the fact that with two components a richer
mechanisms of scattering might occur. When the intensity of $F_2'$ is
further increased (lower panel), still 4 different scattering may
occur but, in this case, it is the scattering between two particles
with $\vect{k}_2$ that induce the instability of the system.

\begin{figure}
\begin{center}
\includegraphics[width=1.0\linewidth,angle=0]{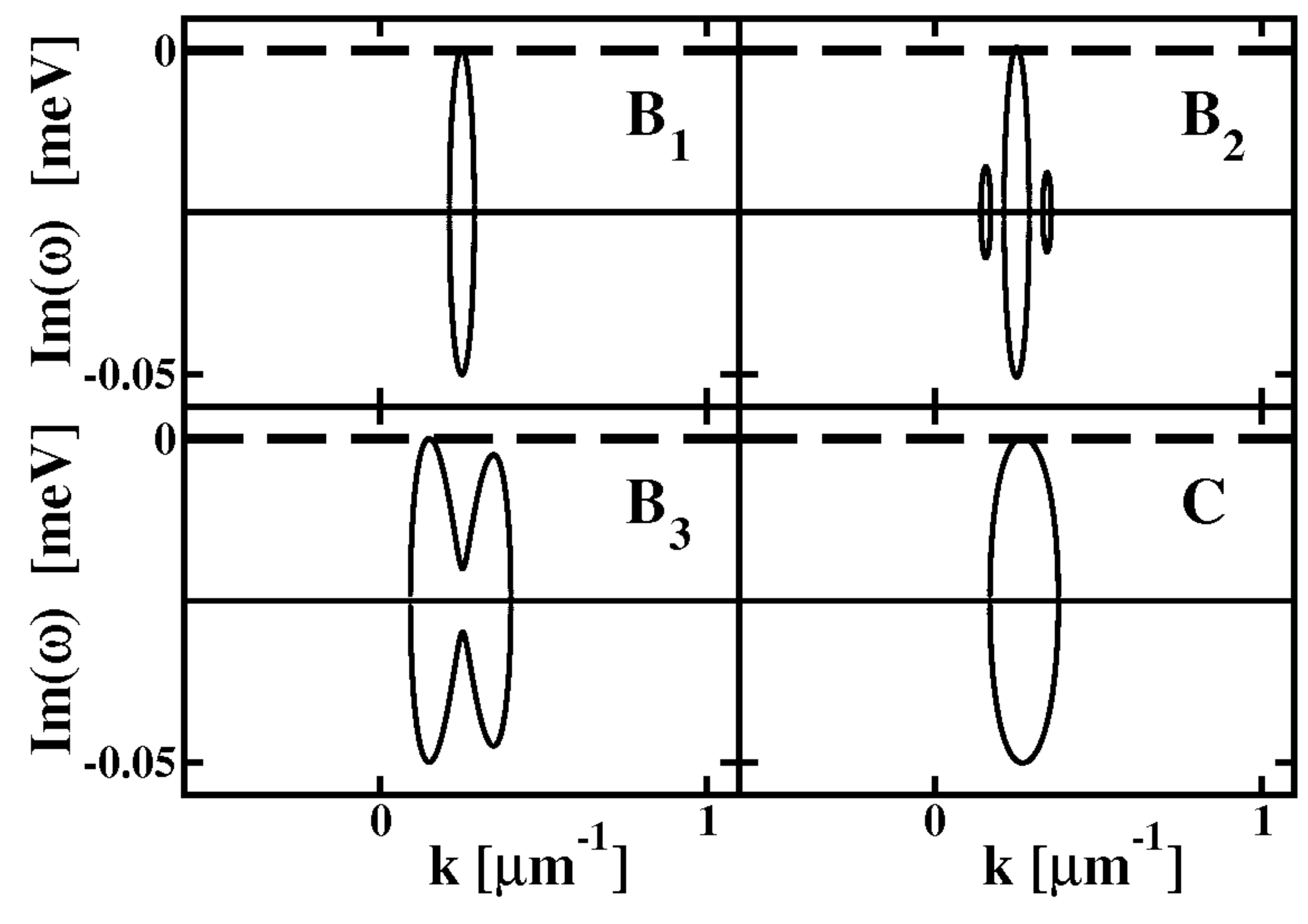}
\end{center}
\caption{Dispersion of the imaginary part of the excitation
  eigenfrequency $\omega=\text{LP}_j^{\pm}$. The three panels
  correspond to points $B_i$ with $i=1,2,3$ and $C$ of
  Fig.~\ref{fig:cuts}, where the higher part of the stability curve
  becomes unstable.}
\label{fig:imaginary2}
\end{figure}

For the transition from stable to unstable regions of the higher
branch of the stability curve we plot the dispersion of the imaginary
part of the excitation eigenfrequency $\omega=\text{LP}_j^{\pm}$ for
points $B_i$ in Fig.~\ref{fig:imaginary2}. In analogy with the case of
fluids with one component, for low intensities of pump 2, the
imaginary part of the dispersion shows a peak at the wavevector of
pump 1, a clear precursor of a Kerr instability. When the intensity of
pump 2 is increased new peaks appear in the imaginary part. In the
case of point $B_2$ two new peaks are precursors of a parametric
instability for the state with $\vect{k}=0.25$ $[\mu m^{-1}]$ even if
the mechanism responsible for the instability of the solution is still
of Kerr type. It is only when the pump intensity $F_2'$ is further
increased (point $B_3$) that the two peaks at $k_-=0.15$ and
$k_+=0.35$ $[\mu m^{-1}]$ became more important and the mechanism of
instability is of the parametric type.  Finally in the lower right
panel (corresponding to point C) of Fig.~\ref{fig:imaginary2} a single
peak at $\vect{k}=\vect{k}_1$ is the precursor of a Kerr-type
instability that ends the region with three stable solutions in the
right panel of Fig.~\ref{fig:cuts}.

\begin{figure}
\begin{center}
\includegraphics[width=1.0\linewidth,angle=0]{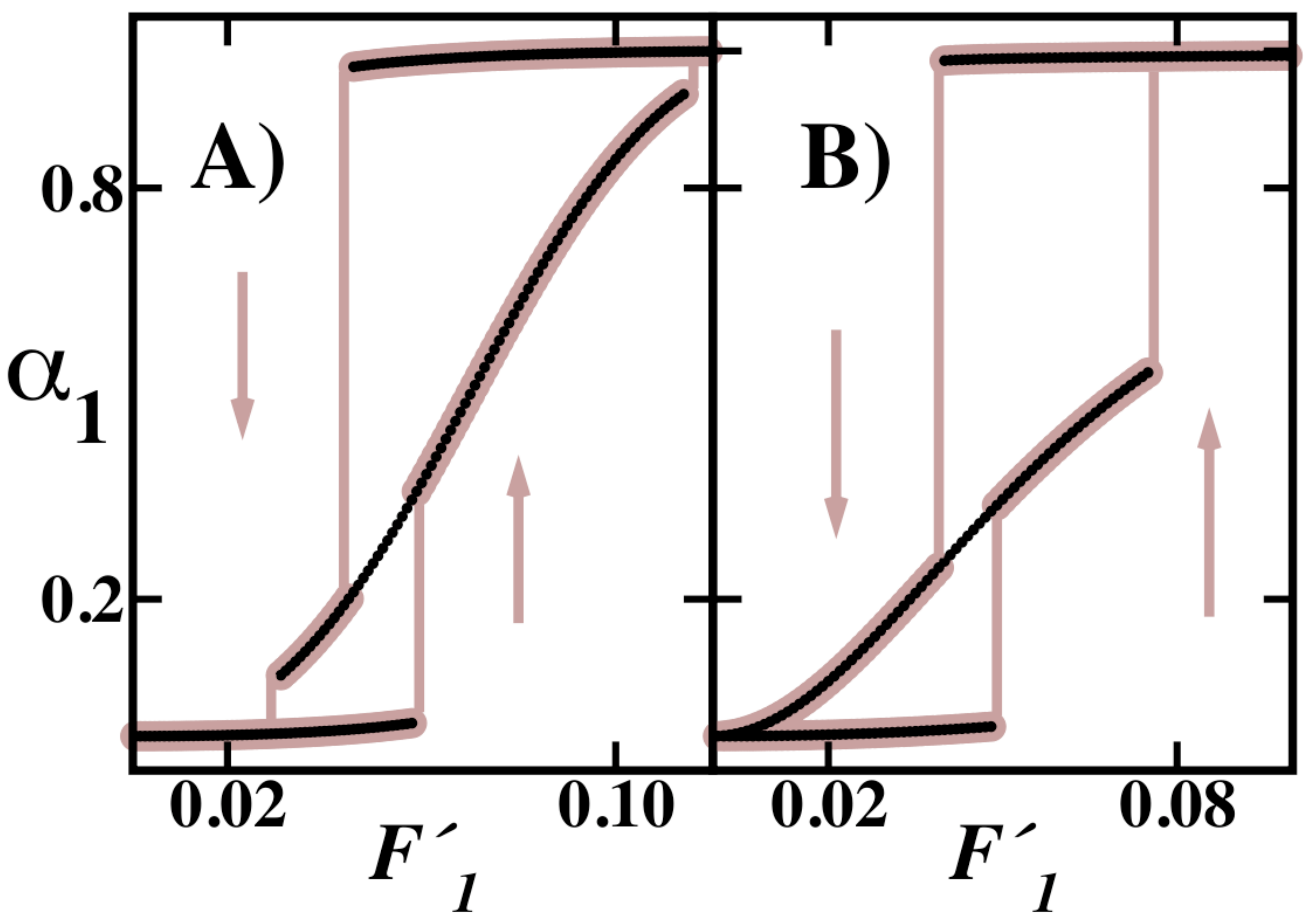}
\end{center}
\caption{\label{fig:hyst}(Color online) Hysteresis cycles of
  $\alpha_1=|\psi_{1_X}^{ss}|^2/(|\psi_{1_X}^{ss}|^2+|\psi_{2_X}^{ss}|^2)$
  [dimensionless] as a function of $F_1'$ for different values of
  $F_2'$ --- stable solutions are in black, while the hysteresis cycle
  performed by the system is in brown. Panel A: $F_2'=0.14 - F_1'$ as
  in the blue line of Fig.~\ref{fig:2d3d}. Panel B: $F_2'= 0.08$
  $[$meV$^{3/2}]$ as in the horizontal black line of
  Fig.~\ref{fig:2d3d}.}
\end{figure}
Multistability also manifests itself in a hysteresis loop for the
populations and emission intensities obtained with a cycle of first
increasing and later decreasing the pumping intensities. Here, the
presence of three stable solutions gives more complicated loops than
the ones obtained for a bistable system in a single-component
polariton fluid. In order to study this aspect, we calculate the
exciton emission intensity at a given frequency $\omega_i$ normalized
to the total exciton emission intensity,
$\alpha_i=|\psi_{i_X}^{ss}|^2/(|\psi_{1_X}^{ss}|^2+|\psi_{2_X}^{ss}|^2)$,
along a closed path of varying pumping intensities. The panels A) and
B) of Fig.~\ref{fig:hyst} respectively show the hysteresis cycles of
$\alpha_1$ when the two pump intensities change along either the blue
or the higher horizontal black line of Fig.~\ref{fig:2d3d}. In panel
A), one starts from a low value of $F_1'$ taking
$F_2'=0.13$~$[$meV$^{3/2}]$ so that the population of state $2$ is
much higher than the population of state $1$, i.e. $\alpha_1 \ll
1$. Increasing $F_1'$ the two populations smoothly evolve until $F_1'
\approx 0.05$ $[$meV$^{3/2}]$ and $F_2' \approx 0.09$
$[$meV$^{3/2}]$. At this point $F_2'$ is too weak to sustain high
population densities in state 2 and, therefore, the system jumps to a
new stable configuration, in which the populations of both states are
low, i.e. $\alpha_1 \approx 0.5$. A further increase of $F_1'$
produces a smooth evolution of the two populations until $F_1'\approx
0.11$ $[$meV$^{3/2}]$ when the system jumps to a third configuration
with a population in state 1 much higher than in state 2,
i.e. $\alpha_1\approx 1$. When we revert the variation of the pumping
intensities along the same path, the jumps to states corresponding to
intermediate and low values of $\alpha_1$ are shifted to the left of
the ones just described for increasing $F_1'$. The multistable
hysteresis loop shown in Fig.~\ref{fig:hyst}A) is related to the fact
that the two pumping lasers are at different pumping angles,
$\vect{k}_i$, and pumping frequencies, $\omega_i$. Therefore the jumps
from low to high population for each component appear at different
values of the pumping intensities, producing the multistable behavior
of $\alpha_1$.

A similar situation occurs when the system evolves along a path on
which one of the pumping intensities remains constant, while the other
varies (black horizontal line at $F_2'=0.08$ of panel 1 in Fig~\ref{fig:2d3d}),
as shown in Fig.~\ref{fig:hyst}B). Starting the with $F_1=0$,
$\alpha_1$ increases smoothly from zero following the lower branch
up to $F_1'=0.06$
 $[$meV$^{3/2}]$. At this point $\alpha_1$ jumps from
values of the order of $0.01$ to $0.3$, corresponding to a population
of state 1 being smaller but non-negligible compared to state 2. As $F_1'$
further increases up to $0.08$ $[$meV$^{3/2}]$, $\alpha_1$ again jumps
abruptly to values of the order of $0.95$. In the reverse process,
$F_1'$ is decreased completing the loop. Also in this case, the jumps
from high to low values of $\alpha_1$ are shifted to the left because
of the different angles and energies at which the lasers are pumping the
two component of the fluid. It is worth noting that the length and the
height of the different plateaus of the hysteresis loops can be efficiently
tuned by carefully choosing the path followed by the intensities, angles
and frequencies of the pumping lasers.

\section{Conclusion}
To summarize, we have studied the stability of a two-component
exciton-polariton fluid under resonant excitation of two pumping
lasers with independently tunable frequencies, angles of incidence and
intensities. We have studied the effect of the detuning between the
laser pump and the bare LP branch, and discussed the different kinds
of instability that might occur for different values of the
parameters. We have shown that, even though the kind of instabilities
are the same as in the one component case (Kerr or parametric), here
the interplay between different instabilities in the two components of
the fluid, can produce a much richer picture.  Moreover, we have shown
that since each component of the fluid jumps between stable states of
its population at a different values of the pumping intensity, the
system sustain multistable hysteresis loops that can be easily
modulated by changing the parameters of the pumping lasers. Finally,
we believe that, due to the wide range of parameters for which the
system is multistable, and due to the increased number of degrees of
freedom with respect to the spin-dependent case, multiatability and
hysteresis loops should be within an experimental reach. Therefore,
the novel system of two-component fluid is a promising candidate for
the realization of optical switches and memories.

%\section{Acknowledgment}
%%%%%%%%%%%%%%%%%%%%%%%%%%%%%%%%%%%%%%%%%%%%%%%%%
%\acknowledgments 
We thank I. Carusotto and D. Whittaker for discussions. This research
has been supported by the Spanish MEC (MAT2008-01555,
QOIT-CSD2006-00019) and CAM (S-2009/ESP-1503). F.M.M. acknowledges
financial support from the program Ram\'on y Cajal.
%%%%%%%%%%%%%%%%%%%%%%%%%%%%%%%%%%%%%%%%%%%%%%%%%

%\bibliography{biblio}

\newcommand\textdot{\.}

\end{document}